\documentclass[preprint2]{aastex}
\usepackage{aastexug}

\def\plotfiddle#1#2#3#4#5#6#7{\centering \leavevmode
\vbox to#2{\rule{0pt}{#2}}
\includegraphics{#1}}

\def\eqalign#1{\null\,\vcenter{\openup\jot
        \ialign{\strut\hfil$\displaystyle{##}$&$
        \displaystyle{{}##}$\hfil \crcr#1\crcr}}\,}

\def\lesssim{\mathrel{\hbox{\rlap{\hbox{\lower4pt\hbox{$\sim$}}}\hbox{$<$}}}}
\def\gtrsim{\mathrel{\hbox{\rlap{\hbox{\lower4pt\hbox{$\sim$}}}\hbox{$>$}}}}
\newcommand{\deltavec}{\mbox{\boldmath $\delta$}}

\newcommand{\uvec}{\mbox{\boldmath $u$}}

\newcommand{\xvec}{\mbox{\boldmath $x$}}
\newcommand{\yvec}{\mbox{\boldmath $y$}}

\begin{document}

\title{Astrometric Detection of Double Gravitational Microlensing Events}

\author{Cheongho Han,}
\affil{Department of Physics, Chungbuk National University,
       Chongju 361-763, Korea}
\email{cheongho@astroph.chungbuk.ac.kr}

\author{Byeong-Gon Park,}
\affil{Bohyeonsan Optical Astronomy Observatory, Korea Astronomy 
        Observatory, Youngchon 770-820, Korea}
\email{bgpark@boao.re.kr}

\author{Wonyong Han}
\affil{Korea Astronomy Observatory, Taejon 305-348, Korea}
\email{whan@kao.re.kr}

\author{Young Woon Kang}
\affil{Department of Earth Sciences, Sejong University, Seoul 143-747, Korea}
\email{kangyw@sejong.ac.kr}

\begin{abstract}
If a gravitational microlensing event is caused by a widely separated 
binary lens and the source approaches both lens components, the source 
flux is successively magnified by the individual lenses: double 
microlensing events.  If events are observed astrometrically, double 
lensing events are expected to occur with an increased frequency due to 
the long range astrometric effect of the companion. We find that although 
the trajectory of the source star image centroid shifts of an astrometric 
double lensing event has a distorted shape from both of the elliptical ones 
induced by the individual single lens components, event duplication
can be readily identified by the characteristic loop in the trajectory 
formed during the source's passage close to the companion.  We determine 
and compare the probabilities of detecting double lensing events from both 
photometric and astrometric lensing observations by deriving analytic 
expressions for the relations between binary lensing parameters to become 
double lensing events.  From this determination, we find that for a given 
set of the binary separation and the mass ratio the astrometric 
probability is roughly an order higher than the photometric probability.
Therefore, we predict that a significant fraction of events that will 
be followed up by using future high precision interferometeric 
instruments will be identified as double lensing events.  
\end{abstract}


\section{Introduction}

One of the most important characteristics of microlensing light curve 
is that it does not repeat.  However, if an event is caused by a widely 
separated binary lens and the source approaches both lens components, 
the source flux is successively magnified by the individual lenses:
double microlensing events \citep{stefano96}.  Due to the special 
geometric condition, however, the chance to become a double lensing 
event is rare.  Therefore, double lensing events have been neglected 
in the previous and current microlensing searches \citep{udalski00, 
alcock00, derue01, bond01}.

Although lensing events have until now been observed only photometrically,
they can also be observed astrometrically by the use of high precision 
interferometric instruments that will be available in the near future,
such as those to be mounted on space-based platforms, e.g.\ the {\it Space 
Interferometry Mission} (SIM) and the {\it Global Astrometric Interferometer 
for Astrophysics} (GAIA), and those to be mounted on 10m class ground-based 
telescopes, e.g.\ Keck and VLT.  If an event is astrometrically observed 
by using these instruments, one can measure the displacement of the source 
star image centroid position with respect to its unlensed position (centroid 
shifts, $\deltavec$).  Once the trajectory of $\deltavec$ is measured, the 
lens mass can be better constrained \citep{miyamoto95, hog95, walker95, 
paczynski98, boden98, han99}.  Recently, astrometric microlensing observation 
is accepted as one of the SIM long term projects (A.\ Gould, private 
communication), and thus astrometric followup observations of events 
detected from the ground-based photometric surveys will become a routine 
process.

One important characteristic of astrometric lensing behavior is that the 
astrometric effect endures to a large lens-source separation where the 
photometric effect is negligible \citep{miralda96}.  Then, it is expected 
that the chance for the source trajectory to enter the astrometrically 
effective lensing region of the companion will be larger, and thus the 
probability of detecting astrometric double lensing events\footnote{We 
define the ``astrometric double lensing event'' as ``the event where the 
path of the source star passes through both of the astrometrically 
effective lensing regions of the individual lens components''.} will 
also be larger.  In this paper, we investigate the general properties 
of astrometric double lensing events and estimate and compare the 
probabilities of detecting double lensing events from both photometric 
and astrometric lensing observations.

The paper is organized as follows.  In \S\ 2, we investigate the general 
properties of double lensing events by presenting and comparing the 
centroid shift trajectories and the light curves of events caused by 
an example wide separation binary.  In \S\ 3, we derive analytic 
expressions for the relations between binary lensing parameters to become 
photometric and astrometric double lensing events and estimate the 
detection probabilities by using these relations.  In \S\ 4, we summarize 
new findings and conclude.

\section{Properties of Double Lensing Events}

The region of photometrically effective lensing region is usually 
expressed in terms of the Einstein ring radius \citep{vietri83, turner84}, 
which is related to the lens parameters by
\begin{equation}
\theta_{\rm E} = \sqrt{4G m\over c^2} 
\left( {1\over D_{\rm ol}} - {1\over D_{\rm os}}\right)^{1/2},
\end{equation}
where $m$ is the lens mass and $D_{\rm ol}$ and $D_{\rm os}$ are the 
distances to the lens and the source, respectively.  If an event is 
caused by a binary where the projected separation between the lens 
components is significantly larger than the sum of the Einstein ring 
radii of the individual lens components, the resulting light curve is 
approximated by the superposition of those of the events where the 
individual lens components work as independent single lenses, i.e.\
\begin{equation}
A \sim A_1 + A_2 -1.
\end{equation}
Here $A_i$ represents the magnification of each single lens event and the 
subscripts $i=1$ and 2 are used to denote the quantities involved with 
the primary\footnote{Here the primary denotes the binary component that 
causes earlier magnification of the source flux and the companion is 
reserved for the other component causing later magnification.} and the 
companion, respectively.  Each single lens event light curve is related 
to the lensing parameters by
\begin{equation}
A_i = {u_i^2+2 \over u_i \sqrt{u_i^2+4}};\qquad 
\uvec_i = \left( {t-t_{0,i}\over t_{{\rm E},i}}\right)\ \hat{\xvec} 
\ +\  \beta_i\ \hat{\yvec}, 
\end{equation}
where $\uvec_i$ represents the separation vector between the source and 
the lens normalized by the Einstein ring radius of the related lens,
$\theta_{{\rm E},i}$, $t_{{\rm E},i}$ is the time required for the 
source to cross $\theta_{{\rm E},i}$ (Einstein time scales), $t_{0,i}$ 
is the time of the source's closest approach to each lens, and $\beta_i$ 
is the lens-source separation (normalized by $\theta_{{\rm E},i}$) at 
that moment (impact parameter).  The units vectors $\hat{\xvec}$ and 
$\hat{\yvec}$ are parallel with and normal to the lens-source transverse 
motion.  By entering the Einstein ring (i.e.\ $u_i\leq 1.0$), the source 
flux is magnified by $A_i\geq 3/\sqrt{5}\sim 1.34$.

\begin{figure*}[t]
\plotfiddle{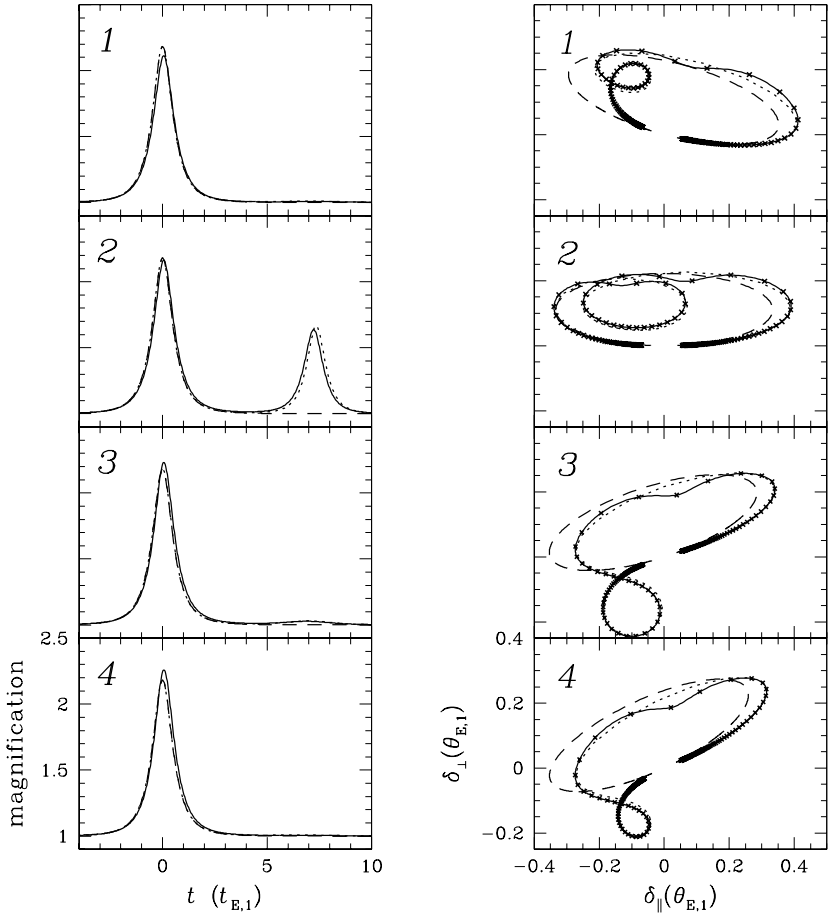}{0.0cm}{0}{70}{70}{-220}{-460}
\vskip12.0cm
\caption{
Example light curves and centroid shift trajectories of events caused by 
a widely separated binary lens system.  Each panel contains three curves, 
where the solid curve is that of the exact binary lens event, the dotted 
curve is that obtained by the single lensing superposition, and the dashed 
curve is that expected without the presence of the companion.  The lens 
system geometry responsible for each event is presented in Fig.\ 2.  The 
number in each panel corresponds to the source trajectory number also 
marked in Fig.\ 2.  Time and length are expressed in units of $t_{\rm E,1}$ 
and $\theta_{\rm E,1}$, i.e.\ the Einstein time scale and the Einstein 
ring radius of the lens causing earlier magnification of the source flux. 
$\delta_\parallel$ and $\delta_\perp$ represent the components of 
$\deltavec$ that are parallel with and normal to the binary axis.
}
\end{figure*}

\begin{figure}[t]
\plotfiddle{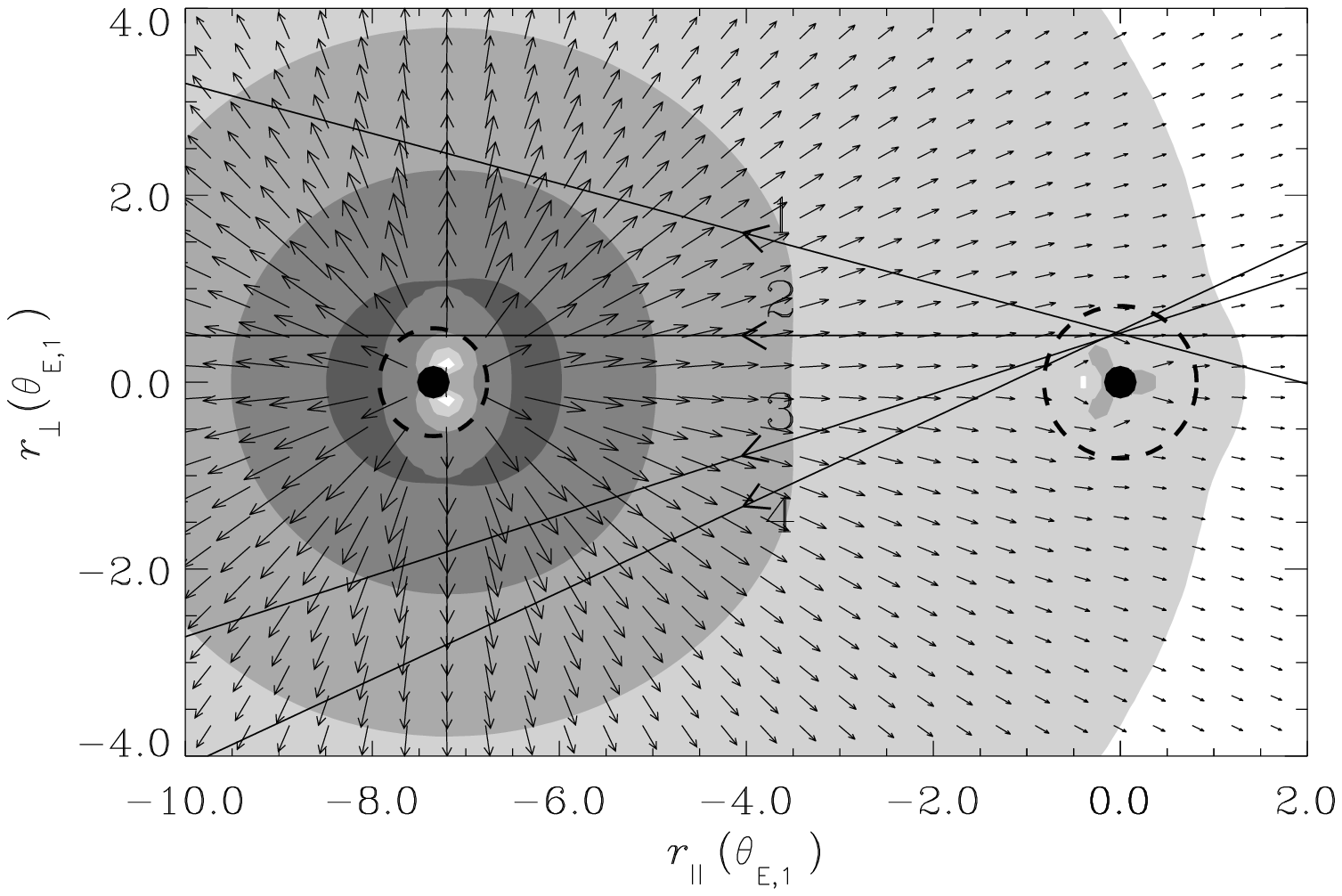}{0.0cm}{0}{55}{55}{-123}{-164}
\vskip5.4cm
\caption{
The lens system geometry responsible for the events whose light curves
and centroid shift trajectories are presented in Fig.\ 1.  The lens
system has a mass ratio between the components of $q=0.5$ and they are
separated by $d=7.35$ in units of $\theta_{\rm E,1}$.  The straight 
lines with arrows represent the source trajectories.  The lenses are 
marked by filled dots and the primary is located at the origin and the 
companion is to the left.  The dashed circles represent the Einstein 
rings of the individual lenses.  The vectors represent the excess 
centroid shifts $\Delta\deltavec$ and the grey-scales are used to show 
the regions where the excesses are greater than $\Delta\delta = 5\%$, 
10\%, 15\% and 20\% of $\theta_{\rm E,1}$, respectively.
}
\end{figure}

Similar to the light curve, the centroid shift of the wide binary event 
is approximated by the superposition of the centroid shift vectors 
induced by the individual single lenses \citep{an01}, i.e.,
\begin{equation}
\deltavec \sim \deltavec_1 + \deltavec_2,
\end{equation}
where the individual centroid shift vectors are represented by
\begin{equation}
\deltavec_i = {\uvec_i \over u_i^2+2}\theta_{{\rm E},i}.
\end{equation}
For a single lens event, the centroid shift follows an elliptical trajectory 
(astrometric ellipse), whose shape (eccentricity) is determined by the 
impact parameter and the size (semi-major axis) is directly proportional 
to the angular Einstein ring radius \citep{walker95, jeong99}.

To show the properties of astrometric double lensing events, in Figure 1, 
we present several example centroid shift trajectories of events caused by 
a widely separated binary lens system.  To compare with the photometric 
lensing behavior, we also present the light curves for the corresponding 
events.  In each panel, we present three different curves,  where the 
solid curve is that of the exact binary lens event, the dotted curve is 
that obtained by the single lensing superposition, and the dashed curve 
is that expected without the presence of the companion.  The lens system 
has a mass ratio between the components of $q=m_2/m_1=0.5$ and they are 
separated by $d = 7.35$ in units of $\theta_{\rm E,1}$.  The source 
trajectory responsible for each event is marked in Figure 2.  To show 
the astrometrically effective region of the companion, we also present
the excess centroid shift vectors of the binary lens system from the 
centroid shifts induced by the primary lens alone, i.e.\ $\Delta\deltavec 
= \deltavec-\deltavec_1$.  Greyscales are used to show the regions where 
the amount of excess is greater than $\Delta\delta = 5\%$, 10\%, 15\% 
and 20\% of $\theta_{\rm E,1}$, respectively.  We note that time and 
lengths are expressed in units of $t_{\rm E,1}$ and $\theta_{\rm E,1}$ 
because the event involved with the primary lens will be the standard 
of all measurements.  To show the changing rate of the centroid position, 
we mark the centroid positions (`{\tt x}' symbol) measured with a time 
interval of $t_{\rm E,1}/4$.  For an event with $t_{\rm E,1}/4$ is a 
month, therefore, this time interval corresponds to a week.

From Fig.\ 1 and 2, one finds the following three important characteristics 
of astrometric double lensing events.
\begin{enumerate}
\item
First, unlike the light curve which is composed of two well separated
light curves induced by the individual lens components, the centroid shift 
trajectory is not simply composed of two separated astrometric ellipses, 
but has a distorted shape from both of the elliptical trajectories.
This is because the centroid shift results from the vector sum of those 
induced by the individual lenses, while the magnification results from 
the scalar sum of the individual magnifications.
\item
Second, despite the distorted shape of the centroid shift trajectory,
event duplication can be readily identified by the characteristic 
loop in the trajectory of $\deltavec$ formed during the source's passage 
close to the companion.  
\item
Third, due to the long range astrometric effect of the companion, as 
expected, astrometric detection of the event duplication will be possible 
even when the separation between the source trajectory and the companion 
is considerable.  By contrast, photometric detection of a double lensing 
event will be possible only when the source approaches very close to the 
companion.
\end{enumerate}

\section{Probabilities of Double Lensing Events}
In the previous section, we have investigated the general properties 
of astrometric double lensing events.  In this section, we derive analytic 
expressions for the relations between the binary lensing parameters to 
become photometric and astrometric double lensing events.  By using these 
relations, we then determine and compare the probabilities of detecting 
double lensing events from both photometric and astrometric lensing 
observations.

\begin{figure}[t]
\plotfiddle{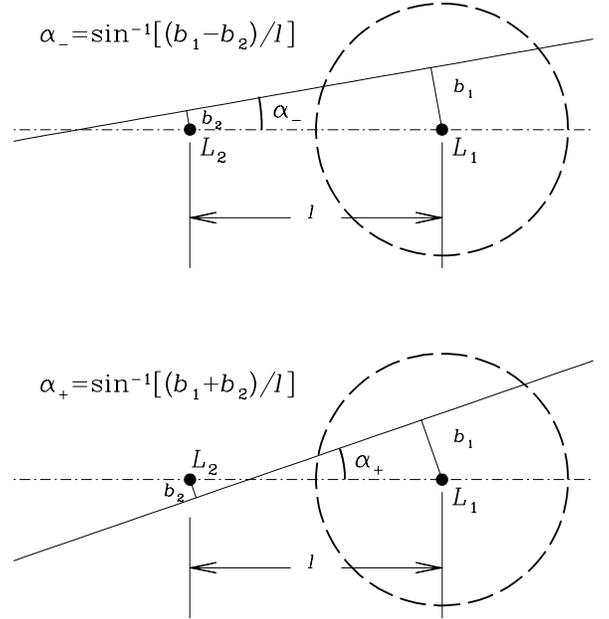}{0.0cm}{0}{52}{52}{-158}{-340}
\vskip8.5cm
\caption{
Geometry of double lensing events.  The binary lens components are marked 
by filled dots and the primary is to the right.  The values $b_1$
and $b_2$ represent the smallest separations between the source trajectory 
to the individual lenses, which are separated by $\ell$. The angle 
$\alpha_\pm$ is the orientation angle of the source trajectory with
respect to the binary axis.  The upper panel is for the case when the 
closest points on the source trajectory from the individual lenses are 
on the same side of the binary axis and the lower panel is when the 
points are located on the opposite sides.
}
\end{figure}

\subsection{Photometric Double Lensing Events}
If $b_1$ and $b_2$ represent the smallest separations from the source 
trajectory to the individual lens components, where projected separation 
between them is $\ell\gg \theta_{\rm E,1}+\theta_{\rm E,2}$, the orientation 
angle of the source trajectory with respect to the binary axis is 
represented by
\begin{equation}
\alpha_\pm = {\rm sin}^{-1}\hskip-1pt \left( {b_1\pm b_2 \over \ell}\right)
\end{equation}
\citep{stefano99}.  Here the sign ``$-$'' is for the case when 
the the closest points on the source trajectory from the individual lenses 
are located on the same side of the binary axis and the ``$+$'' sign is 
when the points are located on the opposite sides (see Figure 3).  When 
all lengths are normalized in units of  $\theta_{\rm E,1}$, equation (6) 
is expressed by
\begin{equation}
\alpha_\pm = {\rm sin}^{-1}\hskip-1pt \left( 
{\beta_1\pm \sqrt{q}\beta_2 \over d} \right),
\end{equation}
where $d=\ell/\theta_{\rm E,1}$.  Let us define $\beta_{\rm th}$ as the 
threshold impact parameter to the companion such that among events for 
which the light variation induced by the primary is detected, only a 
fraction of events whose source trajectories approaching the companion 
closer than $\beta_{\rm th}$ will become double lensing events.  Then, 
under the definition of a photometric double lensing event as ``the 
event where the source trajectory enters both of the Einstein ring radii 
of the individual binary lens components'', the condition to become a 
photometric double lensing event is  $\beta_2 \leq \beta_{\rm th}=1.0$,
and thus the detection probability is computed by
\begin{equation}
P_{\rm ph} = 
{\alpha_{{\rm ph},+} - \alpha_{{\rm ph},-} \over \pi},
\end{equation}
where 
\begin{equation}
\alpha_{{\rm ph},\pm} = 
{\rm sin}^{-1}\hskip-1pt \left( {\beta_1 \pm \sqrt{q}\over d}\right).
\end{equation}

\subsection{Astrometric Double Lensing Events}
Astrometrically, lensing has a longer range effect than the photometric 
effect, and thus a new definition of the threshold impact parameter is 
required.  Let us define $\delta_{\rm th}$ as the threshold amount of the 
centroid shift induced by the companion that is required for an event to 
be identified as an astrometric double lensing event.  Then, the threshold 
impact parameter corresponding to $\delta_{\rm th}$ is obtained by solving 
equation (5) with respect to the lens-source separation, resulting in 
\begin{equation}
\beta_{\rm th} = {1\over 2} \left( {1\over \delta_{\rm th}/\theta_{{\rm E},2}} 
+ \sqrt{{1\over (\delta_{\rm th}/\theta_{{\rm E},2})^2} - 8}\right).
\end{equation}
Note that equation (5) is a quadratic equation of $u$ and thus solving 
the equation results in two values of $\beta_{\rm th}$.  The two solutions 
correspond respectively to the lens-source separations that are smaller 
and larger than $u=\sqrt{2}$, at which $\delta_2$ becomes maximum.  Since 
we are interested only in the maximum allowed lens-source separation for 
the event to be identified as a double lensing event, we take the larger 
value.  If we define a detectable astrometric double lensing event as 
``an event with centroid shift induced by the companion larger than $f$ 
fraction of the Einstein ring radius of the primary lens'', i.e.\ 
$\delta_{\rm th}/\theta_{{\rm E},2}=f\theta_{\rm E,1}/\theta_{\rm E,2}
=f/\sqrt{q}$, equation (10) is expressed in terms of $f$ by
\begin{equation}
\beta_{\rm th} = {1\over 2} \left( {\sqrt{q}\over f} 
+ \sqrt{{q\over f^2} - 8}\right).
\end{equation}
Then the probability of detecting astrometric double lensing event is 
represented by
\begin{equation}
P_{\rm ast} = 
{\alpha_{{\rm ast},+} - \alpha_{{\rm ast},-} \over \pi},
\end{equation}
where 
\begin{equation}
\eqalign{
 & \alpha_{{\rm ast},+} = {\rm sin}^{-1}\hskip-1pt \left({\beta_1 +
  \sqrt{q}\beta_{\rm th}\over d} \right),\cr
 & \alpha_{{\rm ast},-} = {\rm sin}^{-1}\hskip-1pt \left({\left\vert\beta_1 -
\sqrt{q}\beta_{\rm th}\right\vert \over d} \right),\cr
}
\end{equation}
for $d > \beta_1 + \sqrt{q} \beta_{\rm th}$,
\begin{equation}
\alpha_{{\rm ast},+} = {\pi \over 2},\qquad
\alpha_{{\rm ast},-} = {\rm sin}^{-1}\hskip-1pt \left({\left\vert\beta_1 -
\sqrt{q}\beta_{\rm th}\right\vert \over d} \right),
\end{equation}
for $\left\vert\beta_1-\sqrt{q}\beta_{\rm th}\right\vert< d \leq 
\beta_1+\sqrt{q}\beta_{\rm th}$, and
\begin{equation}
\alpha_{{\rm ast},\pm} = \pm{\pi \over 2},
\end{equation}
for $d \leq \left\vert\beta_1 - \sqrt{q} \beta_{\rm th}\right\vert$.  
We note that since the astrometrically effective lensing region of the 
companion can be large enough to extend to or even beyond the position 
of the primary, the threshold orientation angles $\alpha_{{\rm ast},\pm}$ 
have different sets of values depending on the relative size of the 
astrometrically effective region of the companion to the binary separation.

\begin{figure}[t]
\plotfiddle{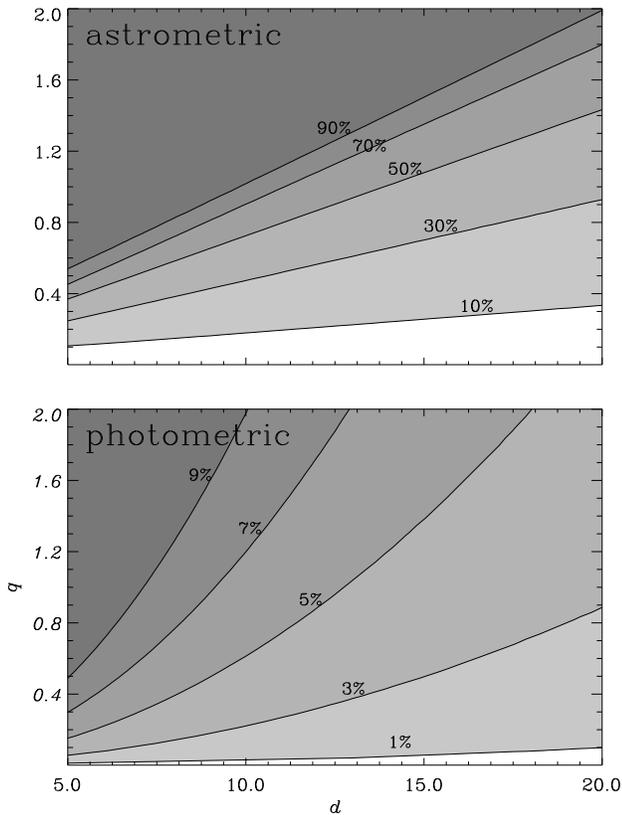}{0.0cm}{0}{74}{74}{-166}{-312}
\vskip10.5cm
\caption{
Comparison of the probabilities of detecting double lensing events from
astrometric and photometric lensing observations.  Note that since the
lens causing later magnification (companion) can be heavier than the
lens causing earlier magnification (primary), the mass ratio can be
larger than $q=1.0$.
}
\end{figure}

In Figure 4, we present the probabilities of detecting astrometric and 
photometric double lensing events as contour maps in the parameter space 
of the binary separation (in units of $\theta_{\rm E,1}$) and the mass 
ratio, $(d,q)$.  For the computation $P_{\rm ast}$, we set $f=0.1$.  If 
the mass and the location of the primary lens are $m=0.3\ M_\odot$ and 
$D_{\rm ol} =4\ {\rm kpc}$, the Einstein ring radius is $\theta_{{\rm E},1}
\sim 500$ $\mu$-arcsec, and thus the imposed threshold centroid shift 
induced by the companion corresponds to $\delta_{\rm th}\sim 50$ $\mu$-arcsec.
We note that the astrometric precision of the SIM will be as low as 
several $\mu$-arcsec, and thus this amount of shift can be easily detected.
Since astrometric followup will be performed only for events where the 
source flux variation is identified by the photometric surveys, we compute 
the probabilities by setting the range of the impact parameter to the 
primary lens to be $0< \beta_1 \leq 1.0$, and the presented probabilities 
are the mean values.  Note that since the lens causing later magnification 
(companion) can be heavier than the lens causing earlier magnification 
(primary), the mass ratio can be larger than $q=1.0$.  To locate the 
unlensed source position, which is the reference position of $\deltavec$ 
measurements, astrometric followup observations of each event will be 
performed for a long period of time, and thus event duplication can be 
identified for events caused by a considerably wide separation binary.  
For an event with $t_{{\rm E},1}\sim 20$ days, the time required to 
identify the duplication caused by a companion with $d\sim 20$ will be 
slightly more than a year.  From the figure, one finds that for a given 
set of $d$ and $q$, the astrometric probability is roughly an order 
higher than the photometric probability.

\section{Conclusion}
We have investigated the properties of double lensing events expected 
to be identified from future astrometric lensing observations.  From 
this investigation, we find that although the centroid shift trajectory 
of an astrometric double lensing event has a distorted shape from those 
of the elliptical ones induced by the individual lens components, the 
event duplication can be readily identified from the characteristic 
loop formed during the source's approach close to the companion.  We 
have also determined and compared the probabilities of detecting double 
lensing events from both photometric and astrometric lensing observations 
by deriving analytic expressions for the relations between the binary 
lensing parameters to become double lensing events.  From this determination, 
we find that for a given set of the binary separation and the mass ratio, 
the probability to detect astrometric double lensing events is roughly an 
order high than the probability to detect photometric double lensing events.
Therefore, we predict that a significant fraction of events that will be 
followed up by using future high precision interferometeric instruments 
will be identified as double lensing events.

\acknowledgements
This work was supported by the National Research Laboratory (NRL)
fund.



\begin{thebibliography}{}

\bibitem[Alcock et al.(2000)]{alcock00}  
	 Alcock C., et al.\  2000, \apj, 542, 281
\vspace{-\abovedisplayskip} \smallskip

\bibitem[An \& Han(2001)]{an01}  
	 An J.\ H., \& Han C.\ 2001, \mnras, submitted
\vspace{-\abovedisplayskip} \smallskip

\bibitem[Boden, Shao \& Van Buren(1998)]{boden98}  
	 Boden A.\ F., Shao M., \& Van Buren D.\ 1998, \apj, 502, 538
\vspace{-\abovedisplayskip} \smallskip

\bibitem[Bond et al.(2001)]{bond01}
         Bond I., et al.\ 2001, \mnras, 327, 868
\vspace{-\abovedisplayskip} \smallskip

\bibitem[Derue et al.(2001)]{derue01}  
	 Derue F., et al.\ 2001, \aap, 373, 126
\vspace{-\abovedisplayskip} \smallskip

\bibitem[Di Stefano \& Mao(1996)]{stefano96}
	 Di Stefano R., \& Mao S.\ 1996, \apj, 457, 93
\vspace{-\abovedisplayskip} \smallskip

\bibitem[Di Stefano \& Scalzo(1999)]{stefano99}  
	 Di Stefano R., \& Scalzo R.\ A.\ 1999, \apj, 512, 579
\vspace{-\abovedisplayskip} \smallskip

\bibitem[Jeong, Han \& Park(1999)]{jeong99}  
	 Jeong Y., Han C., \& Park S.-H.\ 1999, \apj, 511, 569
\vspace{-\abovedisplayskip} \smallskip

\bibitem[Han \& Chang(1999)]{han99}  
	 Han C. \& Chang K.\ 1999, \mnras, 304, 845
\vspace{-\abovedisplayskip} \smallskip

\bibitem[Hog, Novikov \& Polanarev(1995)]{hog95}  
	 H\o g E., Novikov I.\ D., \& Polanarev A.\ G.\ 1995, \aap, 294, 287
\vspace{-\abovedisplayskip} \smallskip

\bibitem[Miralda-Escud\'e(1996)]{miralda96} 
	 Miralda-Escud\'e J.\ 1996, \apj, 470, L113
\vspace{-\abovedisplayskip} \smallskip

\bibitem[Miyamoto \& Yoshii(1995)]{miyamoto95} 
	 Miyamoto M., \& Yoshii Y.\ 1995, \aj, 110, 1427
\vspace{-\abovedisplayskip} \smallskip

\bibitem[Paczy\'nski(1998)]{paczynski98} 
	 Paczy\'nski B.\ 1998, \apj, 404, L23
\vspace{-\abovedisplayskip} \smallskip

\bibitem[Turner, Ostriker \& Gott(1984)]{turner84}
	 Turner E.\ L., Ostriker J.\ P., \& Gott J.\ R.\ 1984, \apj, 284, 1
\vspace{-\abovedisplayskip} \smallskip

\bibitem[Udalski et al.(2000)]{udalski00} 
	 Udalski A., Zebrun K., Szymanski M., Kubiak M., Pietrzynski G.,
         Soszynski I., \& Wozniak P.\ 2000, Acta Astron., 50, 1

\vspace{-\abovedisplayskip} \smallskip

\bibitem[Vietri \& Ostriker(1983)]{vietri83} 
	 Vietri M., \& Ostriker J.\ P.\ 1983, \apj, 267, 488
\vspace{-\abovedisplayskip} \smallskip

\bibitem[Walker(1995)]{walker95} 
	 Walker M.\ A.\ 1995, \apj, 453, 37

\end{thebibliography}
\end{document}